\documentclass{anabs}
\usepackage{graphicx}
\usepackage{times}
\newcommand{\ltsima}{$\; \buildrel < \over \sim \;$}
\newcommand{\simlt}{\lower.5ex\hbox{\ltsima}}
\newcommand{\gtsima}{$\; \buildrel > \over \sim \;$}
\newcommand{\simgt}{\lower.5ex\hbox{\gtsima}}

\newcommand{\pn}{\par\noindent}

\def\lesssim{\mathrel{\hbox{\rlap{\hbox{\lower4pt\hbox{$\sim$}}}\hbox{$<$}}}}
\def\gtrsim{\mathrel{\hbox{\rlap{\hbox{\lower4pt\hbox{$\sim$}}}\hbox{$>$}}}}

\def\aox{$\alpha_{\rm ox}$}

\def\ab1450{$AB_{1450(1+z)}$}

\def\chandra{{\it Chandra\/}}

\def\heao1{{\it HEAO-1\/}}

\Volume{1-2}             
\Year{2003}              
\Month{02}               
\Pagespan{000}{000}      

\begin{document}
\lhead[\thepage]{A.N. Author: Title}
\rhead[Astron. Nachr./AN~{\bf 324} (2003) 1/2]{\thepage}
\headnote{Astron. Nachr./AN {\bf 324} (2003) 1/2, 000--000}

\title{X-rays from the High-Redshift Universe: 
The \textbf{\textit{Chandra}} view}
\author{C. Vignali,\inst{1} W.N. Brandt,\inst{1} D.P. Schneider,\inst{1} 
G.P. Garmire,\inst{1} S. Kaspi,\inst{2} \\ F.E. Bauer,\inst{1} 
\and D.M. Alexander\inst{1}}
%
%
\institute{
Dept. of Astronomy \& Astrophysics, 
The Pennsylvania State University, 525 Davey Lab., University Park, 
PA 16802, USA
\and
School of Physics and Astronomy, 
Tel-Aviv University, Tel-Aviv 69978, Israel}

\correspondence{chris@astro.psu.edu}

\maketitle

\vglue -1.2cm
\section{Introduction}
\vglue -0.2cm

The study of the X-ray properties of $z>4$ quasars has grown substantially 
over the last few years, mostly thanks to \chandra\ 
(e.g., Vignali et al. 2001, 2002; hereafter V01, V02). 
Here we focus on (1) the average X-ray properties of an optically 
selected, luminous sample of nine $z>4$ Palomar Sky Survey (PSS) 
quasars recently observed by \chandra, 
and (2) the \hbox{X-ray} spectral properties of three $z>4$ AGNs in the 
\chandra\ Deep Field-North (CDF-N) using the 2\,Ms exposure.

\vglue -0.5cm
\section{The PSS sample}
\vglue -0.2cm

This sample includes nine quasars \hbox{($z=$~4.09--4.51)} observed 
by \chandra\ and selected from among the optically brightest $z>4$ 
PSS quasars known (V02).
\pn $\bullet$ 
Their broad-band spectral energy distributions are characterized, 
on average, by steeper \aox\ values 
($\langle\alpha_{\rm ox}\rangle$=$-$1.81$\pm{0.03}$) 
than those of lower-luminosity, lower-redshift samples of quasars 
(e.g., the Bright Quasar Survey quasars at \hbox{$z<0.5$} have 
$\langle\alpha_{\rm ox}\rangle$=$-$1.56$\pm{0.02}$). 
A likely explanation is the \aox--UV luminosity anti-correlation 
(e.g., Vignali, Brandt, \& Schneider 2003). 
For comparison, a slightly flatter value is found for the less luminous 
$z>4$ Sloan Digital Sky Survey quasars observed by \chandra\ 
($\langle\alpha_{\rm ox}\rangle$=$-$1.75$\pm{0.03}$; V02;~V01).
\pn $\bullet$ 
Using all of the \chandra\ $z>4$ quasars, a significant 
\hbox{(99.99\% confidence level)}
correlation between \ab1450\ magnitude and soft X-ray flux is found.
%
\pn $\bullet$
An unabsorbed ($N_{\rm H}$$\simlt$8.8$\times$$10^{21}$~cm$^{-2}$ at the 
90\% confidence level) power-law model with 
\hbox{$\Gamma=2.0\pm{0.2}$} is a reasonable fit to the joint 
\hbox{$\approx$~2--30~keV} rest-frame X-ray spectrum of the nine PSS quasars 
(Fig.~1). 
\begin{figure}
\centering
\resizebox{0.57\hsize}{!}
{\includegraphics[angle=-90]{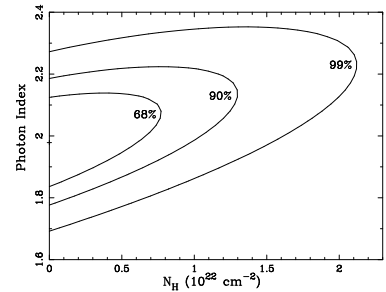}}
\caption{Confidence regions for the photon index 
and intrinsic column density derived from joint spectral fitting of 
the nine PSS quasars.}
\label{fig1}
\vskip -0.2cm
\end{figure}

\vglue -0.5cm
\section{CDF-N AGNs at $\mathbf{z>4}$}
\vglue -0.2cm

Using the 2~Ms exposure, we were able to perform a basic \hbox{X-ray} spectral 
analysis for the three spectroscopically identified $z>4$ AGNs in the CDF-N 
at $z=5.186$, 4.424, and 4.137 (Vignali et al. 2003).
\pn $\bullet$ 
The $z=5.186$ quasar is well fitted with a power law with photon 
index $\Gamma=1.8\pm{0.3}$, consistent with those of lower-redshift, 
unobscured AGNs. 
\pn $\bullet$ 
The other two Seyfert-like AGNs have flatter effective \hbox{X-ray} 
photon indices ($\Gamma\approx$~1.1--1.5), suggesting the presence of 
intrinsic absorption (provided their underlying \hbox{X-ray} continua 
are similar to those of lower-redshift AGNs). 
\pn $\bullet$ 
If the $z=4.137$ AGN suffers from \hbox{X-ray} absorption, the implied 
column density is 
\hbox{$N_{\rm H}\approx$~2$\times10^{23}$~cm$^{-2}$} 
(Fig.~2). 
\vskip -0.5cm
\begin{figure}
\centering
\resizebox{0.60\hsize}{!}
{\includegraphics[angle=-90]{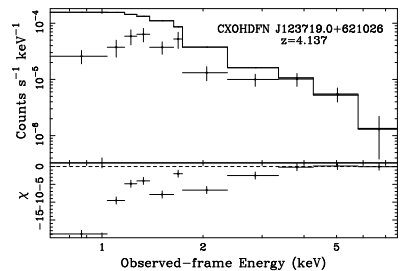}}
\caption{The $z=4.137$ AGN X-ray spectrum 
fitted above 3~keV with a $\Gamma=2$ power-law 
model that has been extrapolated back to lower energies. 
Data-to-model residuals are shown in the bottom panel (in units of $\sigma$).  
A deficit of counts below $\approx$~3~keV is present.} 
\label{fig2}
\vskip -0.6cm
\end{figure}
%

%
\end{document}